\documentclass[showpacs,10pt,twocolumn,prl]{revtex4-1}

\usepackage{amsmath}
\usepackage{amssymb}
\usepackage{graphicx}
\usepackage{amssymb}
\usepackage{graphics}
\usepackage{epsfig}
\usepackage{CJK}
\usepackage{color}

\begin{document}

\title{Magnetic mixed valent semimetal EuZnSb$_2$ with Dirac states in the band structure}
\author{Aifeng Wang$^{1,\dag,*}$, Sviatoslav Baranets$^{2}$, Yu Liu$^{1}$, Xiao Tong,E. Stavitski$^{4}$, Jing Zhang$^{5}$, Yisheng Chai$^{5}$, Wei-Guo Yin$^{1,\P}$, Svilen Bobev$^{2,\S}$ and C. Petrovic$^{1,\ddag}$}
\affiliation{$^{1}$Condensed Matter Physics and Materials Science Department, Brookhaven
National Laboratory, Upton, New York 11973, USA\\
$^{2}$Department of Chemistry and Biochemistry, University of Delaware, Newark, Delaware 19716, U.S.A\\
$^{3}$Center for Functional Nanomaterials, Brookhaven National Laboratory, Upton, NY 11973, USA\\
$^{4}$National Synchrotron Light Source II, Brookhaven National Laboratory, Upton, New York 11973, USA\\
$^{5}$Low Temperature Physics Laboratory, College of Physics, Chongqing University, Chongqing 401331, China}
\date{\today}

\begin{abstract}

We report discovery of new antiferromagnetic semimetal EuZnSb$_2$, obtained and studied in the form of single crystals. Electric resistivity, magnetic susceptibility and heat capacity indicate antiferromagnetic order of Eu with $T_N$ = 20 K. The effective moment of Eu$^{2+}$ inferred from the magnetization and specific heat measurement is 3.5 $\mu_B$, smaller than the theoretical value of Eu$^{2+}$ due to presence of both Eu$^{3+}$ and Eu$^{2+}$. Magnetic field-dependent resistivity measurements suggest dominant quasi two dimensional Fermi surfaces whereas the first-principle calculations point to the presence of Dirac fermions. Therefore, EuZnSb$_2$ could represent the first platform to study the interplay of dynamical charge fluctuations, localized magnetic 4$f$ moments and Dirac states with Sb orbital character.

\end{abstract}

\maketitle

\section{Introduction}

Dirac crystals have attracted great attention in the past years \cite{Wehling}. Topological semimetals with linear energy dispersion in momentum space host variety of quantum transport properties, such as Quantum Hall Effect (QHE), extremely large magnetoresistance (MR) and high mobility \cite{YangK,ShekharC,WangZ}. In particular, manipulation of topologically protected Dirac states by antiferromagnetic (AFM) spins in crystals not only gives rise to rich quantum effects but also enables stable devices that dissipate less energy when compared to traditional FM heterostructures \cite{Smejkal,Shao,ZhangD}.

In contrast to local-moment compounds with $f$-electron levels deep below Fermi energy or heavy fermions that feature magnetic spin exchange with conduction electron bands, mixed i.e. non-integer chemical valence (MV) on 4$f$ or 5$f$ electronic orbitals facilitates charge transfer and absence of long-range magnetic order \cite{Coleman,Fisk}. Charge fluctuations exert considerable influence on the conduction electron bands in general and in particular on the formation of Dirac states at Topological Kondo Insulator surfaces \cite{DzeroM,LuF,DengX,MinCH}. On the other hand, magnetic MV compounds with Dirac topological states are hitherto unknown.

Here we report discovery of a new material, EuZnSb$_2$, with 4$f$ magnetic order below $T_N$ = 20 K. Magnetic moment of Eu is estimated to be 3.5 $\mu_B$, smaller to that of 7.9 $\mu_B$ in EuMnBi$_2$ due to the presence of MV Eu$^{2+}$ and Eu$^{3+}$ atoms, similar to SmB$_{6}$ \cite{MinCH}. Angular-  and temperature-dependent magnetoresistance (MR) are consistent with quasi-two-dimensional Fermi surface. First-principles calculations reveal that square lattice Sb 5$p$$_{x}$/5$p$$_{y}$ - derived bands have a Dirac point between $\Gamma$ and $M$ points in the Brillouen zone. This indicates both the topological electronic bands and magnetic texture. Such arrangement in a tunable ABX$_{2}$ (A = alkaline earth and/or rare earth, B = transition metal) crystallographic structure where graphene-like two-dimensional (2D) quantum charge transport and high mobility were found \cite{KefengSr,KefengCa,ParkJ,LeeG,HuangS,ChenRY,LiuJY2}  can lead to the emergence of rich quantum phases \cite{XuY,Smejkal}.

\section{Experimental and Theoretical Methods}

Single crystals of EuZnSb$_2$ were grown by the Bridgeman method. First, Eu chunks, Zn particles, and Sb lumps were mixed in a stoichiometric ratio, placed into alumina crucible and then sealed in a fused silica tube. The quartz tube was slowly heated to 1030 $^{\circ}$C. After 2 hours, the sample was fast-cooled in four hours to 850 $^{\circ}$C and then slow-cooled to 400 $^{\circ}$C at a rate of 3 $^{\circ}$C/h. Finally, the quartz tube was cooled to room temperature with the furnace power shut off. Single crystals with size up to 4 $\times$ 2 $\times$ 0.5 mm$^3$ can be cleaved from the melted ingot.

Magnetotransport and heat capacity measurement up to 9 T were measured in a Quantum Design PPMS-9. Resistivity $\rho_{xx}$ was measured by a standard four-probe method. Hall resistivity $\rho_{xy}$ was measured by four terminal technique by switching the polarity of the magnetic filed to eliminate the contribution of $\rho_{xx}$.

Powder X-ray diffraction (XRD) data were taken with Cu K$_{\alpha}$ ($\lambda=0.15418$ nm) radiation of Rigaku Miniflex powder diffractometer. The element analysis was performed using an energy-dispersive X-ray spectroscopy (EDX) in a JEOL LSM-6500 scanning electron microscope, confirming stoichiometric EuZnSb$_2$. The X-ray absorption spectroscopy measurement was performed at 8-ID beamline of NSLS II (BNL). The X-ray absorption near edge structure (XANES) and the extended x-ray absorption fine structure (EXAFS) spectra were processed using the Athena software package. The extracted EXAFS signal, $\chi(k)$, was weighed by $k^2$ to emphasize the high-energy oscillation and then Fourier-transformed in a $k$ range from 2 to 11 {\AA}$^{-1}$ to analyze the data in the $R$ space.

X-ray photoemission spectroscopy (XPS) experiments were carried out in an ultrahigh vacuum (UHV) system with base pressures less than 2$\cdot$10$^{-9}$ Torr equipped a hemispherical electron energy analyzer (SPECS, PHOIBOS 100) and twin anode X-ray source (SPECS, XR50). Al K$_{\alpha}$ (1486.7 eV) radiation was used at 10 kV and 30 mA. The angle between the analyzer and X-ray source is 45$^{\circ}$ and photoelectrons were collected along the sample surface normal.

For accurate crystal structure determination, single crystal intensity data sets were collected at 200 K in Bruker SMART APEX II CCD diffractometer using graphite monochromatized Mo K$_{\alpha}$ radiation ($\lambda = 0.71073$ {\AA}). Approximately a quarter of sphere of reciprocal space data was collected in 2 batch runs at different $\omega$ and $\phi$ angles with an exposure time of 6 sec/frame.  A total of 1415 reflections (2$\theta$$_{max}$$\sim$56$^{\circ}$) were collected, 191 of which were unique (T$_{min}$/T$_{max}$ = 0.170/0.327, R$_{int}$ = 0.066).  The data collection, data reduction and integration, as well as refinement of the cell parameters were carried out using the Bruker-provided programs with applied semi-empirical absorbtion correction \cite{SAINT,SADABS}.  The structure was subsequently solved by direct methods and refined on $F^{2}$ (12 parameters) with the aid of the SHELXL package \cite{SHELXL}.  All atoms were refined with anisotropic displacement parameters with scattering factors (neutral atoms) and absorption coefficients \cite{Tables}. The final Fourier map is featureless with the highest residual density and deepest hole of about 3.5 e$^{-}$/{\AA}$^{3}$, situated 0.9 {\AA} and 1.3 {\AA} away from Sb2 and Sb1, respectively. Full occupancies of all sites were confirmed by refinement of the site occupation factors for all sites, none of which deviated from unity within more than 3 standard deviations.  Note that Zn ($Z$ = 30) and Sb ($Z$ = 51) have sufficiently different X-ray scattering factors \cite{Tables}, which makes them clearly distinguishable and ascertains the established, devoid of positional disorder structural model.

For first-principles band structure calculations, we applied the WIEN2K \cite{Blaha} implementation of the full potential linearized augmented plane wave method in the generalized gradient approximation (GGA) \cite{Perdew} + $U_\mathrm{eff}=6$ eV on the Eu 4$f$ orbitals \cite{Larson06,Johannes05,Kunes05} with the spin-orbit (SO) coupling treated in the second variation method. The basis size was determined by $R_\mathrm{mt}K_\mathrm{max}$ = 7 and the Brillouin zone was sampled with a regular $18\times 18 \times 3$ mesh containing 162 irreducible $k$ points to achieve energy convergence of 1 meV. A 10000 $k$-point mesh was used for the Fermi surface calculations.

\begin{figure}
\centerline{\includegraphics[scale=0.38]{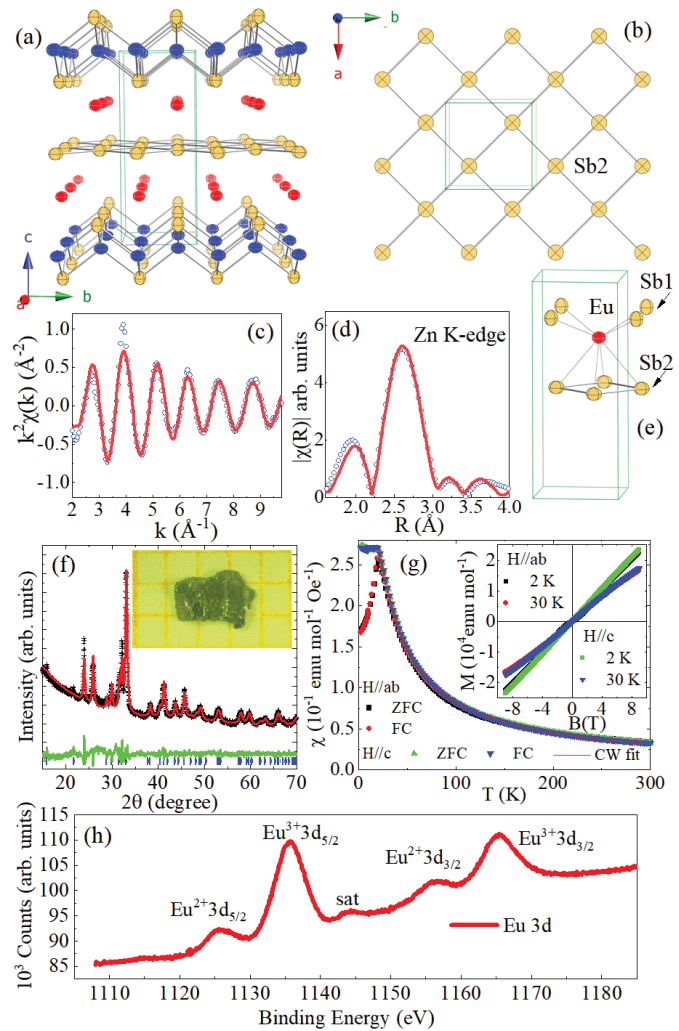}}
\caption{(Color online). (a) Crystal structure of EuZnSb$_2$ projected approximately along [100] with anisotropic displacement parameters drawn at the 90\% probability level and with unit cell outlined. Eu-atoms are depicted as red crossed ellipsoids, Zn atoms are shown with blue ellipsoids, and the Sb1 and Sb2 are shown with yellow ellipsoids, respectively.  (b) Top view of the Sb square nets. EXAFS oscillation with the model fit (c), Fourier transform magnitudes (d) and a close view of the Eu coordination polyhedron (e) at the room temperature. (f) Powder x-ray diffraction pattern of EuZnSb$_2$ confirms phase purity; vertical tick marks denote reflections in the $P4/nmm$ space group. Inset in (f) shows average crystal size on mm scale. (g) Temperature dependent magnetic susceptibility taken in 1 T magnetic field both in the zero field cooling and field cooling mode for H//ab and H//c, respectively. Inset shows field-dependent magnetization at 2 K. (h) XPS core-level spectra of Eu (3$d$) in EuZnSb$_{2}$.}
\label{crystal structure}
\end{figure}

\section{Results and Discussions}

The EuZnSb$_{2}$ refined structure [Fig. 1(a,b), Tables I, II] is ZrCuSiAs type, Pearson index $tP$8. EuZnSb$_{2}$ is a new ternary compound in the Eu-Zn-Sb phase diagram. The other known ternary compounds are EuZn$_{2}$Sb$_{2}$, Eu$_{2}$ZnSb$_{2}$, Eu$_{21}$Zn$_{4}$Sb$_{18}$ and Eu$_{11}$Zn$_{6}$Sb$_{12}$ \cite{EuZn2Sb2,Eu2ZnSb2,Eu21Zn4Sb18,Eu11Zn6Sb12}. EuZnSb$_{2}$ is isotypic to ZrCuSiAs [space group $P$4/$nmm$ (No. 129), $Z$=2]. There are close to 150 compounds of this type, whose structures are confirmed and deposited in structural databases. Of those, only one other pnictide containing an alkaline-earth metal forms with the same structure and it is the bismuthide CaMnBi$_{2}$ \cite{BrechtelE,KefengCa}. No other Eu-based pnictides are known to exist with this structure. This is also true for the EuMnBi$_{2}$ \cite{May} which is structurally similar, but not identical to EuZnSb$_{2}$. Strontium, whose solid state chemistry often mirrors that of Eu does not form pnictide phase isotypic to EuZnSb$_{2}$. A Sr-analog with the 1:1:2 stoichiometric ratio, SrZnSb$_{2}$ is known, but forms with the orthorombic CaMnSb$_{2}$ structure type ($Pnma$, Pearson index $oP$16 \cite{Brechtel}. Importantly, the vast majority of pnictides with the tetragonal ZrCuSiAs structure type are with trivalent rare-earth metals and their structures show a discernable tendency to be transition metal-deficient. A good and relevant example of this notion is the series of RZn$_{1-x}$Sb$_{2}$ (R = La, Ce, Pr, Nd, Sm, Gd, Tb), where the Zn-site can be vacant with amount of defects being as high as 40 \% \cite{Zelinska}. In this regard, EuZnSb$_{2}$ represents a rare compound with the tetragonal ZrCuSiAs/HfCuSi$_{2}$ structure, not only considering its chemical makeup, but also the fact that it is a zinc-pnictide from this family that is apparently as close to being a stoichiometric phase as possible.

The structure prototype is known and will not be described at length. Results of structural refinement are shown in Table I. There are four independent sites in the asymmetric unit (Table II) and based on the refinements, the structure appears to be devoid of disorder on any of them. The nearly spherical shape of the anisotropic displacement parameters is also a testament to this conjecture. The structure is based on PbO-type layers made up of fused ZnSb$_{4}$-tetrahedra and square-nets of Sb atoms [Fig. 1(a-b)]. All Zn-Sb interatomic distances are within the common range (Table II) and within the range for the sum of the single-bonded covalent radii. The Sb-Zn-Sb angles within the ZnSb$_{4}$-tetrahedra deviate from the ideal 109.5$^{\circ}$ value and range from 104.4(2)$^{\circ}$ to 112.1(1)$^{\circ}$. There are no indications from the structure refinements that the Sb square lattice is a subject to a Peierls distortion.  The Sb-Sb interatomic distances within the square nets are longer than what is expected for a single covalent Sb-Sb bond, but the atomic interactions in such topology are hypervalent, consistent with the longer distances [3.105(3) {\AA}].

\bigskip
\begin{table}[tbp]\centering%
\caption{Selected single-crystal data collection and structure refinement parameters for EuZnSb$_{2}$ measured at 200 K using $\lambda$ = 0.71073 ${\AA}$. The corresponding crystallographic information file (CIF) has been deposited with the Cambridge Crystallographic Database Centre (CCDC) - depository number - 1999191}
\begin{tabular}{cc}
\hline\hline
Refined composition & EuZnSb$_{2}$ \\
Formula Mass (g/mol) & 460.83 \\
Space Group & $P$4/$nmm$ (No. 129) \\
Z & 2 \\
$a$ (${\AA}$) & 4.392(4) \\
$c$ (${\AA}$) & 11.21(2) \\
$V$ (${\AA}$$^{3}$) & 216.2(5) \\
Density (g/cm$^{3}$) & 7.08 \\
$R1$ [$I>2\sigma (I)$)]$^{a}$ &
0.058 \\
$wR2$ [$I>2\sigma (I)$)]$^{a}$ & 0.129 \\
$\Delta$$\rho$$_{max,min}$/e$\cdot$${\AA}$$^{-3}$ & 3.39,3.96 \\ \hline\hline
\multicolumn{2}{c}{$^{a}$ $R1=\Sigma |F_{0}|-|F_{c}|/\Sigma |F_{0}|$,} \\
\multicolumn{2}{c}{$wR2=[\Sigma (|F_{0}^{2}-|F_{c}^{2}|)^{2}/\Sigma
(wF_{0}^{2})^{2}]^{1/2}$} \\
\multicolumn{2}{c}{$w$=1/[$\sigma$$^{2}$$F_{0}^{2}$+(0.0714$\cdot$P)$^{2}$+(9.825$\cdot$P)] and P=($F_{0}^{2}$+2$F_{c}^{2}$)/3}
\end{tabular}%
\label{1}%
\end{table}%

\bigskip
\begin{table}[tbp]\centering%
\caption{Selected single-crystal data collection and structure refinement parameters for EuZnSb$_{2}$ measured at 200 K using $\lambda$ = 0.71073 ${\AA}$.}%
\begin{tabular}{cccccc}
\hline\hline
Atom & Site & $x$ & $y$ & $z$ & $U_{eq}$ \\
Eu1 & 2$c$ & 1/4 & 1/4 & 0.2673(2) & 0.015(1) \\
Zn1 & 2$a$ & 3/4 & 1/4 & 0 & 0.020(1) \\
Sb1 & 2$c$ & 1/4 & 1/4 & 0.8479(3) & 0.017(1) \\
Sb2 & 2$b$ & 3/4 & 1/4 & 1/2 & 0.018(1) \\ \hline
\multicolumn{6}{c}{Selected interatomic Distances (${\AA}$)} \\
Zn1 & -Sb1 $\times$ 4 & 2.779(3) & Eu1 & -Sb1 $\times$ 4 & 3.363(3) \\
Sb2 & -Sb2 $\times$ 4 & 3.105(4) &  & -Sb2 $\times$ 4 & 3.409(4) \\
\hline\hline
\end{tabular}%
\label{2}%
\end{table}%

We note that ABX$_{2}$ (A = alkaline earth and/or rare earth, B = transition metal) materials feature rather tunable crystal structure. The square Bi or Sb slabs host graphene-like Dirac states of X=Bi/Sb orbital character with charge- or spin-density-wave order coupled to topological states \cite{KefengLaAgSb2,ZhangA,GuoYF,RahnMC,SakoptaA}. Weyl states were proposed in Sr$_{1-y}$Mn$_{1-z}$Sb$_2$ and YbMnBi$_2$ \cite{LiuJY,Borisenko}. Topological Dirac states were also found in Zn-based 112 materials such as BaZnBi$_2$ and SrZnSb$_2$ \cite{WangKF,ZhaoK}; EuMnBi$_2$ hosts QHE with field tunable Eu 4$f$ magnetic order and magnetopiezoelectric effect \cite{Masuda,ShiomiY}.

Figures 1(c) and 1(d) show the Zn K-edge oscillations and corresponding Fourier transform magnitudes of Extended X-ray absorption fine structure (EXAFS) spectra of EuZnSb$_2$, respectively. In the single-scattering approximation, the EXAFS could be described by the following equation \cite{Prins}:
\begin{align*}
\chi(k) = \sum_i\frac{N_iS_0^2}{kR_i^2}f_i(k,R_i)e^{-\frac{2R_i}{\lambda}}e^{-2k^2\sigma_i^2}sin[2kR_i+\delta_i(k)],
\end{align*}
where $N_i$ is the number of neighbouring atoms at a distance $R_i$ from the photoabsorbing atom. $S_0^2$ is the passive electrons reduction factor, $f_i(k, R_i)$ is the backscattering amplitude, $\lambda$ is the photoelectron mean free path, $\delta_i$ is the phase shift of the photoelectrons, and $\sigma_i^2$ is the correlated Debye-Waller factor measuring the mean square relative displacement of the photoabsorber-backscatter pairs. The corrected main peak around $R \sim 2.77$ {\AA} in Fig. 1(d) corresponds to the Zn-Sb1 bond distances in ZnSb$_{4}$ tetrahedra and is in good agreement with single crystal refinement result (Table II). The Eu atoms are coordinated by eight nearest neighbor antimony atoms in a square-antiprismatic fashion [Figure 1(e)] with Eu-Sb1 and Eu-Sb2 distances longer than 3.3 {\AA} (Table II). Unit cell from the powder X-ray diffraction experiment on pulverized crystals [Fig. 1(f)] can be fitted well with the $P4/nmm$ structural model (Table I), confirming single crystal refinement result and phase purity.

Temperature dependence of magnetic susceptibility measured with zero field cooling and field cooling mode is shown in Fig. 1(g) for two primary crystallographic directions. A sharp peak at 20 K was observed for the magnetic susceptibility with H//$a$, while a saturation behavior is observed below 20 K with H//$c$, indicating that the magnetic easy axis is along the $a$-axis. The peak at 20 K can be explained by the antiferromagnetic transition of Eu$^{2+}$. Magnetic susceptibility above 20 K exhibits a typical Curie-Weiss behavior, which can be well fitted to $\chi=\chi_{0}+\frac{C}{T-\theta}$, where $\chi_0$ is the temperature independent Pauli contribution and $C$ is related to the effective moments. The effective moment of Eu$^{2+}$ is 3.5 $\mu_B$ and $\theta$ = -9.4 K for H//$ab$ while 3.7 $\mu_B$ and $\theta$ = -11.2 K for H//$c$. Negative Curie temperatures indicate dominant antiferromagnetic interactions of Eu$^{2+}$. In order to obtain more information about the magnetic structure of Eu$^{2+}$, we perform the field dependent measurement of magnetization loop, as shown in the inset in Fig. 1(g). The magnetization along the $c$-axis exhibits perfect linear behavior at 2 K and 30 K, i.e. at temperatures below and above $T_N$. When field is applied along the $ab$ plane, the magnetization shows linear behavior in 30 K whereas a very small deviation from linearity is observed at 1.1 T in M(H) taken at 2 K. The temperature and field dependence magnetization of EuZnSb$_2$ [Fig. 1(g)] resembles that of EuZnBi$_2$ whose magnetic structure is analogous to EuMnBi$_2$ spin-flop AFM phase \cite{Masuda,May}. However, these Bi-based materials belong to $I$4/$mmm$ space group containing different Eu-sublattice structure and fixed Eu valence when compared to EuZnSb$_{2}$. The weak in-plane anisotropy of Eu$^{2+}$ gives rise to the kinks around 1.1 T. For EuZnSb$_{2}$ the M vs H do not saturate up to 9 T, and no spin-flop is observed, in contrast to that in EuZn$_2$Sb$_2$ \cite{Weber}. Thus, the magnetic field that might induce the spin-flop of Eu$^{2+}$ can be higher than 20 T  \cite{Masuda}.

X-ray photoelectron 3$d$ core level spectra [Fig. 1(h)] shows spin-orbit split states 3$d_{5/2}^{2+}$, 3$d_{3/2}^{2+}$, 3$d_{5/2}^{3+}$ and 3$d_{3/2}^{3+}$, at binding energies 1125, 1155, 1135  and 1165 eV, respectively, confirming the presence of both Eu$^{2+}$ and Eu$^{3+}$. We also observe a weak satellite peak at somewhat higher binding energies from 3$d_{5/2}^{3+}$, originating from  the multielectronic excitations in the photoelectron emission. The XPS spectra is rather similar to Eu$_{2}$SrBi$_{2}$S$_{4}$F$_{4}$ \cite{Haque} and provide explanation for the observed paramagnetic Curie-Weiss moment reduced from 7.9 $\mu_B$. Interestingly, both Eu$^{2+}$ and Eu$^{3+}$ share the same $2c$ atomic site in the $P4/nmm$ unit cell; this might reduce Zn vacancies when compared to La$_{1-x}$ZnSb$_{2}$ \cite{Sologub}. The absence of Zn defects and mixed valent Eu might suggest that the square-planar nets in EuZnSb$_{2}$ can not be treated as hypervalent Sb since for trivalent rare earth defects on the Zn site are necessary to lower the valence electron count \cite{Sologub,TremelW}.

\begin{figure}
\centerline{\includegraphics[scale=0.32]{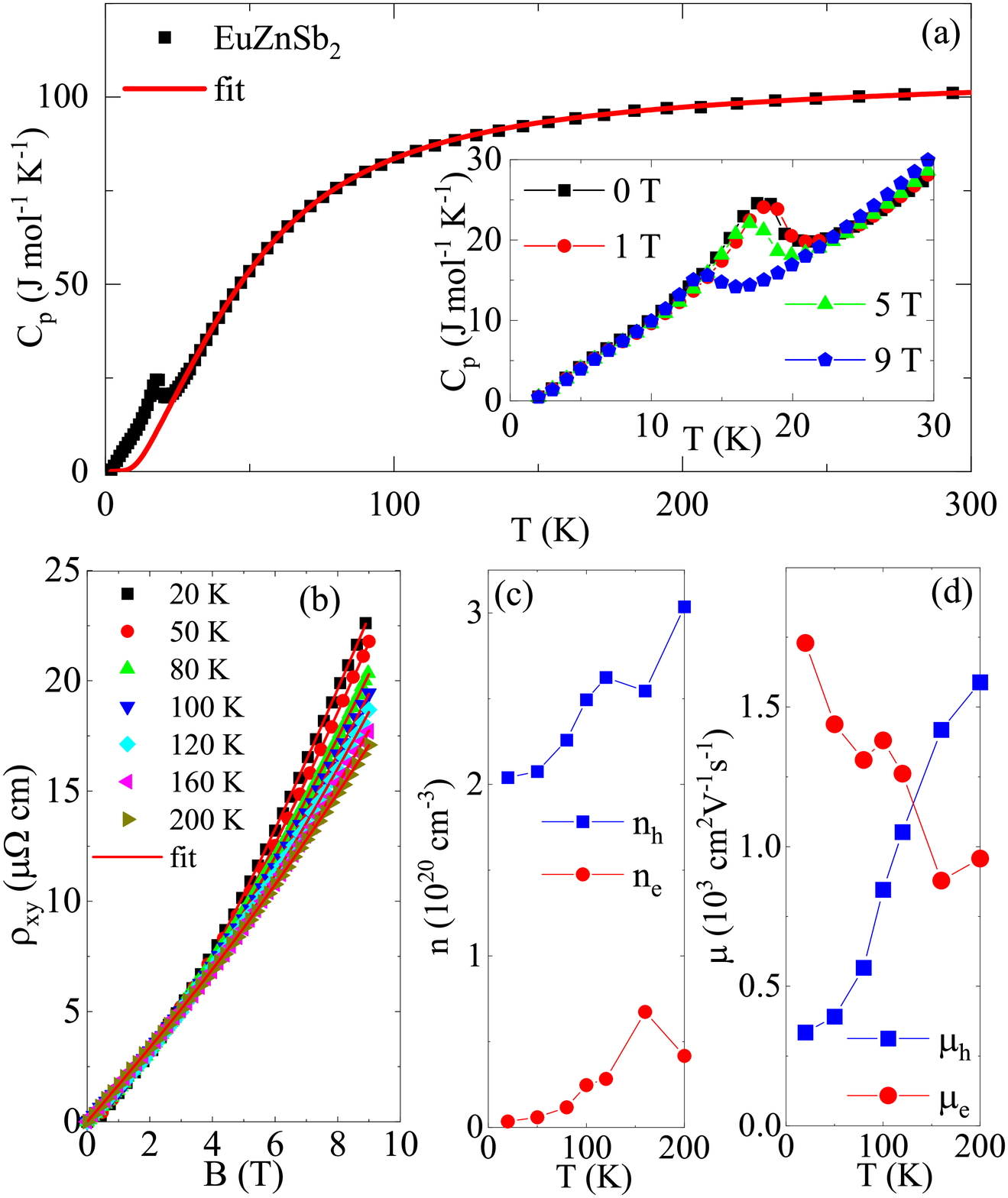}}
\caption{(Color online). (a) Temperature dependence of specific heat for EuZnSb$_2$. The red line represents the fit. Inset shows the low-temperature specific heat in various magnetic field. (b) Magnetic field dependence of Hall resistivity $\rho_{xy}$ at different temperatures; two band model fits (see text) are shown by red lines. Inferred carrier concentration (c) and mobility (d) from the two-band fitting.
}
\end{figure}

Temperature dependence of the specific heat of EuZnSb$_2$ is shown in Fig. 2(a). The anomaly corresponding to the AFM transition of Eu$^{2+}$ is observed below 20 K [Fig. 2(a) inset]. The anomaly is suppressed to low temperature by magnetic field, consistent with resistivity and magnetization measurement.
Room temperature heat capacity is close to 3$NR$, where $N$ is the atomic number per chemical formula and $R$ is the universal gas constant 8.314 J mol$^{-1}$ K$^{-1}$. Specific heat above 25 K is well fitted by the Debye-Einstein model \cite{Prakash}:
$$
\begin{aligned} C_{e l+p h}(T)=& \gamma T+\alpha 9 n R\left(\frac{T}{\theta_{D}}\right)^{3} \int_{0}^{\theta_{D} / T} \frac{x^{4} e^{x}}{\left(e^{x}-1\right)^{2}} d x \\ &+(1-\alpha) 3 n R \frac{\left(\theta_{E} / T\right)^{2} e^{\theta_{E} / T}}{\left(e^{\theta_{E} / T}-1\right)^{2}} \end{aligned}
$$
Where $\theta_D$ and $\theta_E$ are the Debye and Einstein temperatures, respectively, and $\alpha$ denotes the relative contribution of Debye and Einstein terms to phonon heat capacity. The $\theta_D$ = 233 K and $\theta_E$ = 73 K can be obtained from the fitting. We note that inclusion of Einstein term was necessary to fit $C(T)$, suggesting the presence of optical phonon modes \cite{Walti,Fischer}. The entropy change induced by the AFM transition of Eu$^{2+}$ is obtained by $\Delta{S}$ = $\int(C_p - C_{ph+el})/T$$dT$, and $\Delta{S}$ is estimated to be 11 J mol$^{-1}$ K$^{-1}$, smaller than the theoretical value $\Delta S = R \ln (2 J+1)$ = 17.3 J mol$^{-1}$ K$^{-1}$ for Eu$^{2+}$ with $J = \frac{7}{2}$. However, we can infer the $J$ = 1.38 from $\Delta{S}$ = 11 J mol$^{-1}$ K$^{-1}$. Then, the effective moment of Eu$^{2+}$ is estimated to be 3.6 in agreement with the magnetization measurement and also with XPS investigation that reveals the presence of both Eu$^{2+}$ and Eu$^{3+}$.

Hall resistivity $\rho_{xy}$ exhibits positive slope and small temperature dependence [Fig. 2(b)], suggesting dominant hole carriers. Clear nonlinear behavior at high field indicates multiband transport, similar to YbMnSb$_2$ \cite{WangYY}. We fit $\rho_{xy}$ with semiclassical two-band model \cite{Smith}:
$$
\rho_{x y}=\frac{B}{e} \frac{\left(n_{h} \mu_{h}^{2}-n_{e} \mu_{e}^{2}\right)+\left(n_{h}-n_{e}\right)\left(\mu_{h} \mu_{e}\right)^{2} B^{2}}{\left(n_{h} \mu_{h}+n_{e} \mu_{e}\right)^{2}+\left(n_{h}-n_{e}\right)^{2}\left(\mu_{h} \mu_{e}\right)^{2} B^{2}}
$$
where $n_e$ ($n_h$) and $\mu_e$ ($\mu_h$) denote the carrier concentrations and mobilities of electrons and holes, respectively. The obtained carrier concentration and mobility are shown in Fig. 2(c) and 2(d), respectively. The hole concentration is nearly one order of magnitude larger than that of electrons whereas at low temperatures electron-type carriers have much larger mobility, in contrast to values at 200 K.

\begin{figure}
\centerline{\includegraphics[scale=0.38]{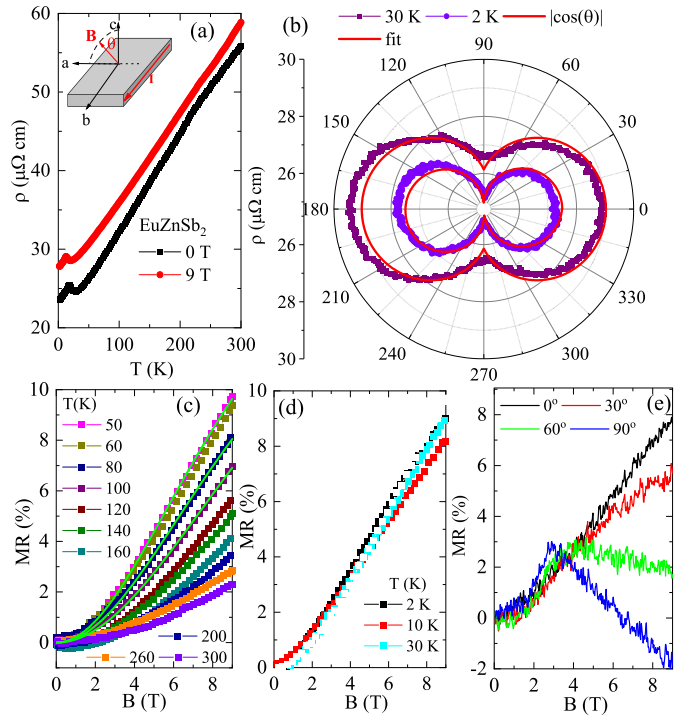}}
\caption{Color online). (a) Temperature dependence of the in-plane resistivity of EuZnSb$_2$ in 0 and 9 T, respectively. (b) Angle $\theta$ dependence of resistivity at 2 and 30 K, respectively measured in 9 T. (c,d) Magnetoresistance of EuZnSb$_{2}$ for $\theta$ = 0. Green lines show fits using two-band orbital magnetoresistance obtained using carrier concentrations and mobilities from Fig. 2(c,d); we note that such fits are possible for carrier concentration and mobility variations for up to $\pm$10\% and $\pm$20\% from values reported in Fig. 2. (e) Magnetoresistance at different tilt angles $\theta$ between $B$ and $c$-axis. Data were measured at 2 K and the current was always perpendicular to the magnetic field.}
\label{crystal structure}
\end{figure}

Resistivity of EuZnSb$_2$ decreases on cooling; anomaly corresponding to the AFM transition of Eu$^{2+}$ is observed at 20 K [Fig. 3(a)]. There is an upward shift of the temperature-dependent resistivity and the $T_{N}$ is suppressed to 12 K in 9 T applied along the $c$-axis [$\theta$ = 0 in Fig. 3(a) inset]. The resistivity is similar to that of AZnSb$_2$ ($A$ = Ba and Sr) and EuZnBi$_2$, but is in contrast to semiconducting EuMnSb$_2$ \cite{RenWJ, LiuJ, YiC}. Since MR is determined by the extremal cross section of the Fermi surface perpendicular to the magnetic field direction, information about geometric structure of Fermi surface can be obtained from the angular dependence of MR. As shown in Fig. 3(b), MR angular dependence is anisotropic. It shows a typical 2D behavior, $B \propto |cos\theta|$ \cite{KefengSr,KefengCa,LiuJ}. Similar angular dependence at 2 K and at 30 K suggests that the magnetic order of Eu$^{2+}$ is only a small perturbation on $\rho(\theta)$.

Magnetic field dependence of resistivity at different temperatures for $\theta$ = 0 is shown in Fig. 3(c,d). The maximum MR ratio, $\mathrm{MR}=\left[\rho(B)-\rho(0)\right]/\rho(0)\times$100\%, is 10 \% at 50 K and 9 T, which is relatively small. MR(B) hints towards a low field $B^2$ dependence to high field linear MR crossover, as is sometimes observed in Dirac materials \cite{KefengSr,KefengCa,RenWJ,LiuJ}. Semiclassical MR in metals features $B^2$ dependence in the low field and a saturating MR in high fields \cite{metal}.

Strong external magnetic field might lead to a complete quantization of the perpendicular orbital motion of carriers with linear energy dispersion and quantized Landau Levels $
E_n=sgn(n)v_F\sqrt{2e\hbar B|n|}$ where $n=0,\pm1,\pm2,\cdots$ is the LL index and $v_F$ is the Fermi velocity \cite{LL1,LL2}. In such quantum limit, the energy difference from the lowest to the $1^{st}$ Landau level is $\triangle_{LL}=\pm v_F\sqrt{2e\hbar B}$; $\triangle_{LL}$ is larger than Fermi energy $E_F$ and the thermal fluctuations $k_BT$ so that all carriers are at the lowest Landau level where linear MR is expected \cite{metal,quantummr}. Crossover magnetic field $B^*$ above which the quantum limit is satisfied at
specific temperature $T$ $B^*=\frac{1}{2e\hbar v_F^2} (E_F+k_BT)^2$ \cite{qt3} could then be used to estimate Fermi velocity.  Such analysis, assuming quantum limit, gives the Fermi velocity $v_F\sim 5.13\times 10^5$ ms$^{-1}$, i.e. $\Delta$$_{LL}$ = 5 meV .  We estimate Fermi energy from $E_F$=$\hbar$$^2$/m$_0$(3$\pi$$^2$$n$)$^{2/3}$ where $n$ is the carrier density. By taking measured carrier density [Fig. 2(c)] we obtain $E_F$ = 16.5 meV for electron pocket and $E_F$ = 250 meV for the hole pocket, suggesting that MR arises from the first few Landau levels. Indeed, the two-band orbital magnetoresistance MR:\cite{SmithR}
$$
MR = \frac{n_e\mu_en_h\mu_h(\mu_e + \mu_h)^2(\mu_0H)^2}{{{{({\mu _e}{n_h} + {\mu _h}{n_e})}^2} + {{({\mu _h}{\mu _e})}^2}{{({\mu _0}H)}^2}{{({n_h} - {n_e})}^2}}}
$$
is satisfactory only at high temperatures [Fig. 3(c)].

Other possible reasons for linear MR in high fields include mobility fluctuations in an inhomogeneous crystal \cite{Parish} or open orbits. The former can be excluded since EuZnSb$_{2}$ is stoichiometric crystal. The latter usually arises for electronic motion on orbits associated with magnetic field oriented along elongated necks of the Fermi surface, for example in Cu \cite{BianQ}, in topological materials with extremely large magnetoresistance with compensated charge carriers \cite{ZhangSN} or in two-dimensional conductors where magnetic field is applied parallel to the conducting layers \cite{KovalevAE} .  Whereas we note that EuZnSb$_{2}$ is not a compensated metal with equal electron and hole concentration [Fig. 2(c)], and that in Fig. 3 (c,d) magnetic field is oriented orthogonal to quasi-2D conduction direction, we can not exclude the possibility for open Fermi surface pockets. It also should be noted that small negative MR appears when magnetic field B is tilted away from $c$-axis at 2 K [Fig. 2(d)]. This likely arises due to in-plane spin reorientation which could indicate shift of the magnetic easy axis with external field rotation.

\begin{figure}
\centerline{\includegraphics[scale=0.7]{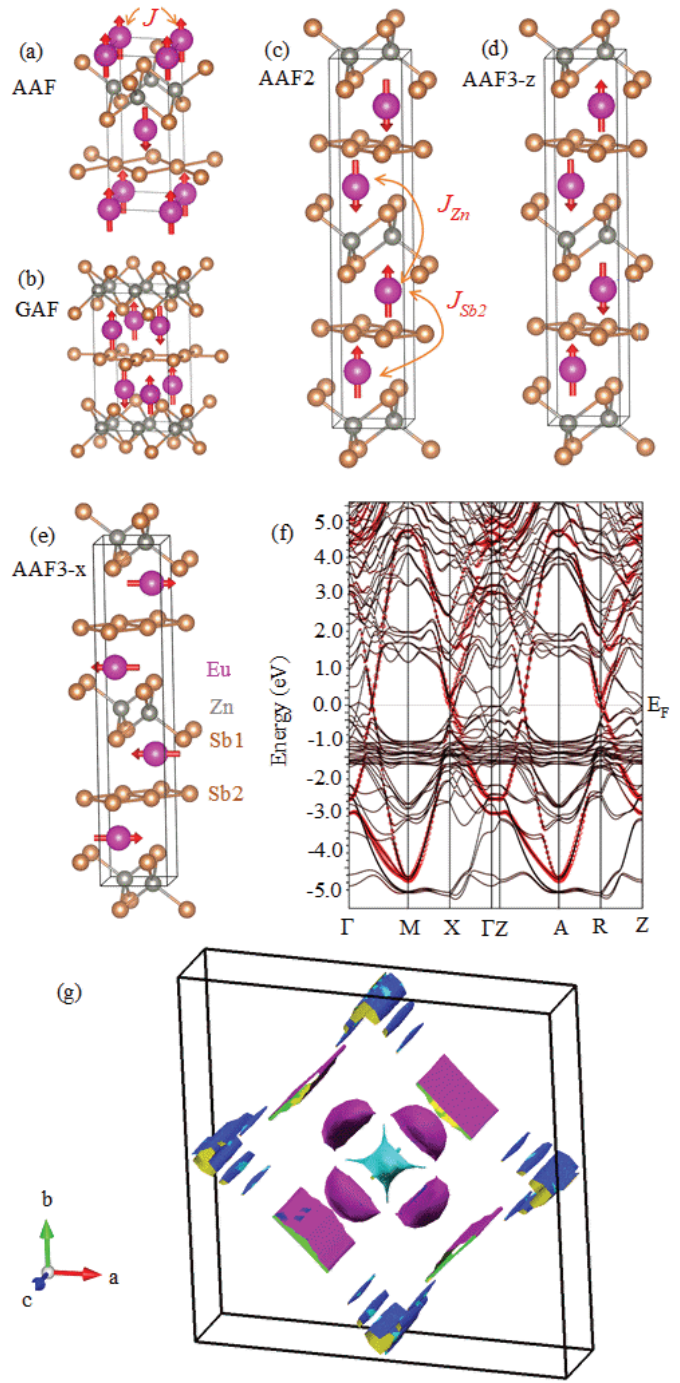}}
\caption{((Color online). (a-e) The antiferromagnetic structures used in the first-principles total energy calculations. (f) The electronic band structure calculated with the AAF3-$x$ structure (e) in GGA+$U$+SO. Red thick circles denote the $5p_x$/$5p_y$ band derived from the Sb2 square lattices. (g) The calculated Fermi surface of EuZnSb$_2$. Warm colors denote hole carriers and cold colors denote electrons. Electron-like bands are located near the $\Gamma$ point and $X$ point, hole bands are along the (0.0)-($\pi$,$\pi$) line excluding the $\Gamma$ point states.
}
\end{figure}

\bigskip
\begin{table}[tbp]\centering%
\caption{The first-principles total energy per formula unit of five different magnetic patterns as shown in Fig. 4. FM denotes the ferromagnetic configuration.
}
\begin{tabular}{cccc}
\hline\hline
Pattern & GGA & GGA+SO & GGA+$U$+SO \\ \hline
AAF    &0         &    0           & \\
GAF    &2.88      &    4.49        & \\
FM     &1.13      &    $-0.53$     & \\
AAF2   &6.46      &    7.24        & \\
AAF3-$z$\footnote{Magnetization along the $z$ axis.} &$-6.23$   &    $-6.18$     & 0  \\
AAF3-$x$\footnote{Magnetization along the $x$ axis.} &          &    $-2.85$     & $-0.04$ \\
\hline\hline
\end{tabular}%
\label{magnetic}
\end{table}%

First-principles band structure calculations reveal the mechanism of Dirac point formation and point to possible magnetic structure after comparing the total energy for five different magnetic structures [Fig. 4(a-e), Table~\ref{magnetic}]. The exchange interactions $J_\mathrm{Sb2}S^2=-1.810 (-1.616)$ meV (antiferromagnetic), $J_\mathrm{Zn}S^2=1.544 (1.558)$ meV, $JS^2=0.856 (0.663)$ meV are estimated from GGA with (without) inclusion of spin-orbit coupling [see Figs.4(a) and (c) for the definition of the exchange paths]. These numbers indicate that the AAF3 pattern is the ground state. Fig.4(e) shows the low-energy magnetic structure with the in-plane easy axis obtained by GGA+$U$+SO calculations with $U=6$ eV for the Eu $4f$ orbitals. However, the state is almost degenerate with the one with the out-of-plane easy axis, meaning that the magnetic anisotropy is very weak. The moments of Eu$^{2+}$ feature ferromagnetic arrangement in plane, and are stacked antiferromagnetic by every two layers in the out plane direction. In its corresponding electronic structure  [Fig.4(f)], the flat bands lying about 1.3 eV below the Fermi level are derived from the Eu 4$f$ orbital. They are moved down from about two hundred meV below the Fermi level in GGA calculations by inclusion of $U$, a feature that is more severe in the DFT calculations for Eu$^{3+}$N~\cite{Johannes05} but nearly absent for Eu$^{2+}$ chalcogenides~\cite{Larson06}, suggesting that the Eu ionicity in EuZnSb$_2$ be in between EuN and EuO. The square-lattice Sb $5p_x$/$5p_y$ derived bands have a Dirac point between the $\Gamma$ and $M$ points, indicating that the system contains both the topological electronic bands and the magnetic texture. The Fermi surface [Fig.4(g)] shows a coronavirus-like electron pocket at $\Gamma$ point surrounded by four hole pockets. The Dirac states around ($\pi$/2, $\pi$/2) are flat along the $c$-axis, i.e. they are restricted to the Sb2 square lattice. Overall, the volume of the hole pockets are larger than the electron pocket, in agreement with the experiments. The structure at the $\Gamma$ point features elongated necks which could contribute to open orbits and linear MR, as discussed above. The antiferromagnetic $J_\mathrm{Sb2}$ (i.e., the exchange interaction between two nearest Eu ions above and below the square-lattice Sb2 layer) means that the time reversal symmetry is not broken for the itinerant electrons on the Sb2 layers at the level of the mean-field-like calculations, leading to the four-fold degenerated Dirac states. Yet, it would have been broken by the random distribution of Eu$^{2+}$ and Eu$^{3+}$ ions on the same crystal $2c$ site---leading to the formation of Weyl states---which cannot be described by standard DFT calculations with the $1\times 1 \times 2$ minimal structure model. Further studies of topological electronic structure and putative Weyl states in EuZnSb$_2$ using angular resolved photoemission (ARPES) and large-supercell DFT calculations are of high interest.

\section{Conclusion}

In conclusion, we report discovery of EuZnSb$_{2}$, a new magnetic semimetal. EuZnSb$_{2}$ exhibits MV Eu, i.e. both Eu$^{2+}$ and Eu$^{3+}$ are present in the unit cell. The Eu$^{2+}$ moments order antiferomagnetically below 20 K. First-principle calculations are consistent with the presence of Dirac states in the band structure whereas magnetotransport suggests dominant quasi-2D Fermi surface sheets. Further ARPES and neutron experiments are of interest to shed more light on Dirac dispersion in momentum space and magnetic space group in the ordered state. Moreover, spectroscopic studies are of interest to investigate possible coupling of dynamical charge fluctuations and the long-range magnetic order with Dirac bands.

Work at Brookhaven is supported by the U.S. DOE under Contract No. DE-SC0012704. Cryogenic magnetization measurements at Chongqing were supported by Fundamental Research Funds for the Central Universities (2018CDJDWL0011) and Projects of President Foundation of Chongqing University (2019CDXZWL002). Single crystal diffraction carried out at the University of Delaware was supported by the U.S. Department of Energy, Office of Science, Basic Energy Sciences, under Award No. DE-SC0008885. This research used resources of the Center for Functional Nanomaterials, which is a U.S. DOE Office of Science Facility, at Brookhaven National Laboratory under Contract No. DE-SC0012704.

$^{*}$Present address: School of Physics, Chongqing University, Chongqing 400044, China
$^{\dag}$afwang@cqu.edu.cn
$^{\P}$ wyin@bnl.gov
$^{\S}$ bobev@udel.edu
$^{\ddag}$petrovic@bnl.gov

\end{document}